\documentclass[pra,epsfig,showpacs,preprint]{revtex4}
\usepackage[dvips]{graphicx}

\begin{document}

\title{Distinguishing coherent atomic processes using wave mixing}

\author{
    A.~M.~Akulshin,
    and R.~J.~McLean
}

\affiliation{ Centre for Atom Optics and Ultrafast Spectroscopy,
Swinburne University of Technology, Melbourne, Australia 3122
\\}


\begin{abstract}

We are able to clearly distinguish the processes responsible for enhanced low-intensity atomic Kerr nonlinearity, namely coherent population trapping and coherent population oscillations in experiments performed on the Rb D1 line, where one or the other process dominates under appropriate conditions. The potential of this new approach based on wave mixing for probing coherent atomic media is discussed. It allows the new spectral components to be detected with sub-kHz resolution, which is well below the laser linewidth limit. Spatial selectivity and enhanced sensitivity make this method useful for testing dilute cold atomic samples.

\end{abstract}


\maketitle

Promising applications in quantum information processing and quantum metrology as well as fundamental aspects of atom-light interaction are generating widespread interest in coherent atomic media \cite{ARIMONDO96,BUDKER02,MARANGOS05,BOOK09}. Designing of atomic media to have predetermined properties requires proper consideration of effects related to hyperfine and Zeeman coherent population trapping (CPT) \cite{ARIMONDO96} as well as coherent population oscillations (CPO) \cite{BOYD-CPO} both of which contribute to a variety of low-intensity nonlinear processes occurring in atomic media. CPT and CPO exhibit quite different spectral and spatial properties but for a range of experimental conditions in real atomic media their contributions interfere making interpretation of spectra difficult. In this report we investigate the interplay of these effects in Rb vapours excited by resonant bi-chromatic laser radiation on the D1 absorption line and show that the wave mixing approach allows a clear separation of the two effects.

Long-lived anisotropy due to Zeeman coherence and light-induced CPO both depend crucially on the polarizations of the applied radiation.  We generate these effects with two monochromatic, mutually coherent optical fields at frequencies  $\nu_{1}$ and $\nu_{2}$ and use the optical heterodyne method to analyze the RF spectra of the beat signals for parallel and orthogonal polarizations of the applied laser radiation.  In essence, applied fields at $\nu_{1}$ and $\nu_{2}$  having parallel polarizations can generate efficient population oscillations at $\delta ={\nu_{2}-\nu_{1}} \leq\Gamma/2\pi$ between the ground and excited levels, where $\Gamma/2\pi\simeq$ 6 MHz is the natural width of the Rb D1 line, while for orthogonal polarizations Raman-type transitions between different magnetic sublevels are favored and the excitation of CPO is not straightforward.

\begin{figure}[b] 
\includegraphics[angle=0, width=10cm]{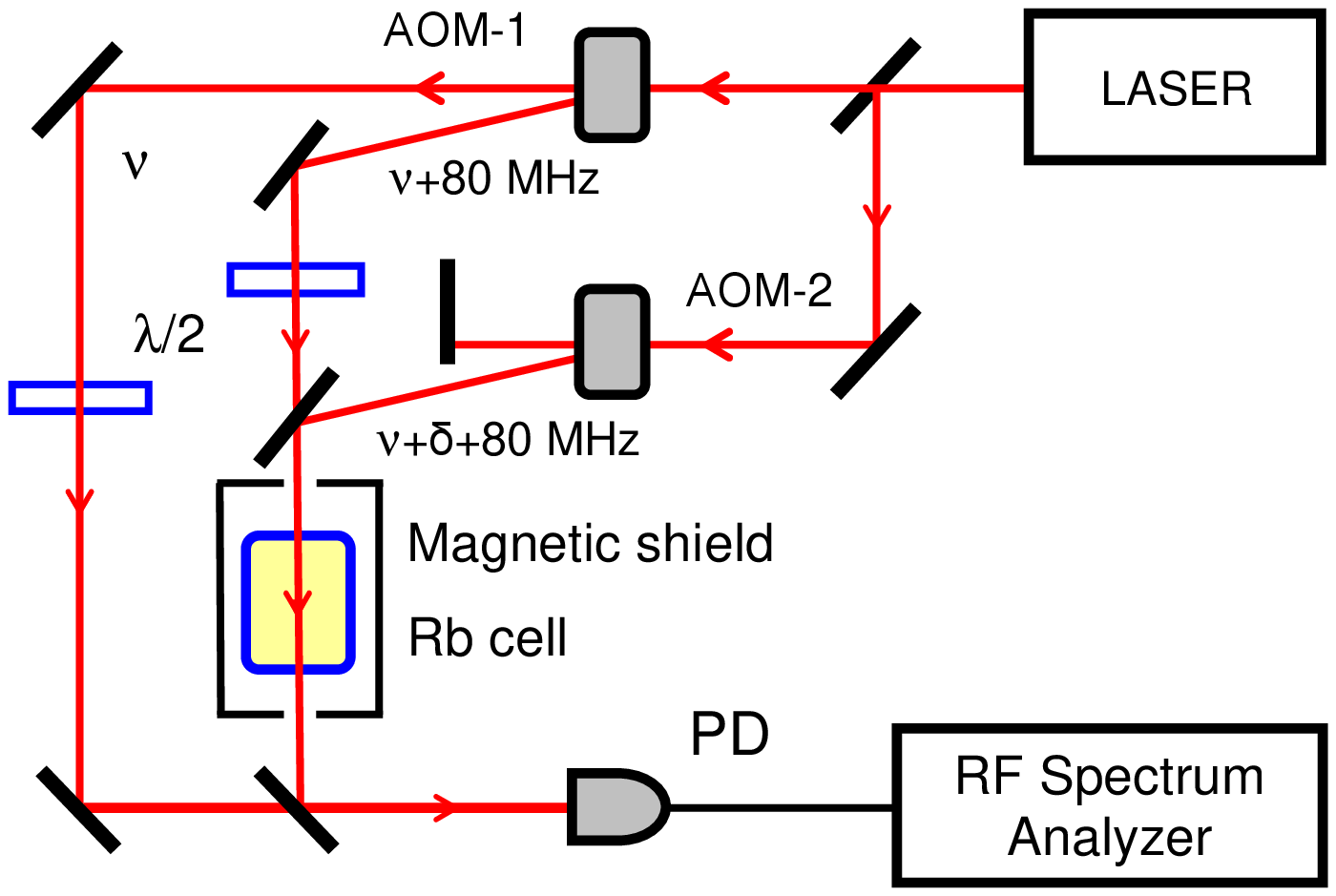}
\caption{Experimental setup.
   \label{fig.1}
   }
\end{figure}

Figure 1 shows the scheme of the experimental setup. An extended-cavity diode laser with approximately 500 kHz linewidth is tuned to the Rb D1 absorption line at 795 nm.  Two mutually coherent optical fields with tunable frequency offset $\delta$ are produced with two acousto-optic modulators (AOMs) driven by fixed frequency (80 MHz) and tunable ($\delta$ + 80 MHz)  oscillators.  The  diffracted beams are carefully combined on a 50/50 non-polarizing beam splitter and sent through a 5-cm long glass cell that contains a natural mixture of Rb isotopes with no buffer gas. Thus, Rb atoms are excited by co-propagating laser beams in the vicinity of the Raman resonance between degenerate magnetic sublevels of the same ground state hyperfine level. As the coherent optical heterodyning technique requires a reference field that is mutually coherent with the applied radiation and frequency detuned from the Raman resonance, the frequency-unshifted beam transmitted through AOM-1 is employed for this purpose.  This beam is combined with the radiation transmitted through the cell. Subsequent mixing on a fast photodiode produces beat signals that are observed and processed on an RF spectrum analyzer. The typical power of each optical beam applied to the Rb cell is $\simeq$ 1 mW, while the power of the reference beam is 2 mW. The cross section of the beams in the cell is approximately 2 mm$^2$. The polarizations of the beams are controlled via the polarizers and wave plates.  Doppler-free saturated absorption resonances obtained in an auxiliary Rb cell not shown in Fig. 1 are used as frequency references for the laser.

The cell temperature is set within the range 30-60 °C by a magnetic-field-free heater. A $\mu$-metal shield reduces ambient magnetic fields to the level of a few mG in the cell.

In the case of orthogonal polarizations, which are more frequently considered for CPT studies, the applied radiation at $\nu_{1}$ and $\nu_{2}$, if resonant with the $^{85}Rb$  $5S_{1/2}(F_g=3)\rightarrow5P_{1/2}(F_e=2)$ transition for example, couples all magnetic sublevels of the F=3 ground state hyperfine level by two-photon ($\pi-\sigma^{\pm}$)-type Raman transitions.  This can generate a high-contrast electromagnetically induced transparency (EIT) resonance and high nonlinear susceptibility $\chi^{(3)}$ in the vicinity of the Raman resonance \cite{SCHMIDT96,AKU-JOB-04}.  High nonlinearity of the atomic medium is a necessary condition for efficient wave mixing, but angular momentum conservation must also be satisfied. The conditions for nearly degenerate four wave mixing (FWM) in coherent Rb vapour at sub-mW power level have been discussed in \cite{AKU-JPB-11}.

\begin{figure}[t] 
\includegraphics[angle=0, width=14cm]{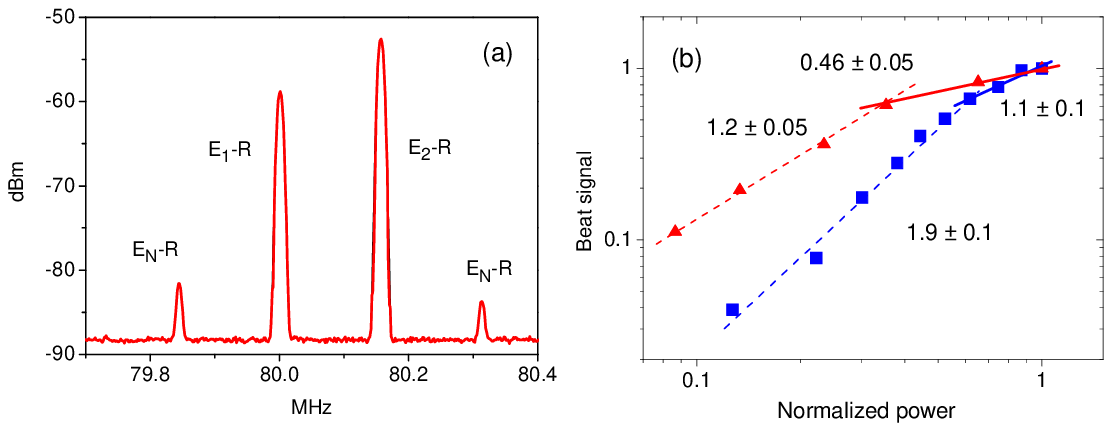}
\caption{a) RF beat signal for orthogonal linear polarizations of the applied laser components tuned to the $^{85}$Rb $5S_{1/2}(F_g=3)\rightarrow5P_{1/2}(F_e=2)$ transition. The linear polarization of the reference field is at 45$^0$ in respect to the polarization of the transmitted light. $\delta \simeq$ 175 kHz; atomic density $N \simeq 6\times 10^{10} cm^{-3}$. $I_{1,2}\simeq25~ mW/cm^2$; b) Low-frequency $E_N$-$R$ beat signal vs. normalized power of the optical fields at $\nu_1$ (blue squares) and $\nu_2$ (red triangles); $\delta \simeq$ 700 kHz. Numbers show the slopes of linear fits to the experimental points.
   \label{fig.2}
   }
\end{figure}

Figure 2a shows a beat spectrum obtained with orthogonal linear polarizations of the applied laser light on the $^{85}$Rb $5S_{1/2}(F_g=3)\rightarrow5P_{1/2}(F_e=2)$ transition.  The peaks labeled $E_1$-$R$ and $E_2$-$R$ are due to mixing of the reference field and the applied radiation with electric field amplitudes $E_1$ and $E_2$, respectively.  While the $E_1$-$R$ and $E_2$-$R$ peaks are present for both resonant and non-resonant laser light, the appearance of the $E_N$-$R$ peaks is evidence of Stokes and anti-Stokes field generation due to atom-light interaction. The absorption of two photons from the field with  amplitude $E_1$ and frequency $\nu_1$ and stimulated emission at frequency $\nu_2$ of the component with amplitude $E_2$ gives rise to an optical wave with electric field amplitude $E_N \sim\chi^{(3)}E_1^2E_2\sim\chi^{(3)}P_1(P_2)^{1/2}$, where $P_1$ and $P_2$ are the power of the applied fields.
The new wave at $\nu_N=2\nu_1 - \nu_2=\nu_1-\delta$  when mixed with the reference field on the photodiode  produces a beat signal at 79.825 MHz, with amplitude proportional to $\chi^{(3)}E_1^2E_2E_R\sim \chi^{(3)}P_1(P_2P_R)^{1/2}$, where $E_R$ and $P_R$ are the electric field amplitude and power of the reference field.

\begin{figure}[t] 
\includegraphics[angle=0, width=14cm]{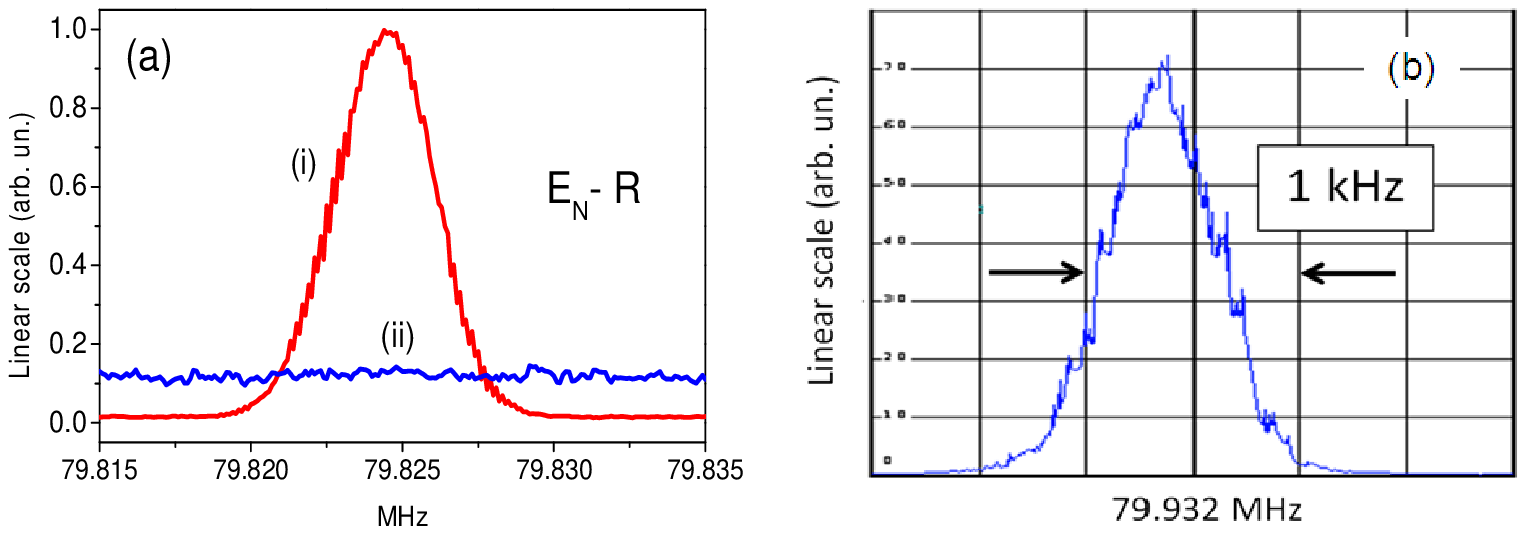}
\caption{a) Curves (i) and (ii) represent the $E_N$-$R$ beat signals for the $5S_{1/2}(F_g=3)\rightarrow5P_{1/2}(F_e=2)$ transition and $5S_{1/2}(F_g=2)\rightarrow5P_{1/2}(F_e=3)$ transitions, respectively. Curve (ii) is plotted with ten times enhanced sensitivity. b) Sub-kHz wide beat signal between the reference and the new field at $\nu_1-\delta$ obtained for the case of orthogonal linear polarization on the $5S_{1/2}(F_g=3)\rightarrow5P_{1/2}(F_e =2)$ transition; $\delta \simeq$ 68 kHz.
   \label{fig.3}
   }
\end{figure}

Figure 2b shows dependence of the low-frequency $E_N$-$R$ peak on the power of the applied fields $P_1$ and $P_2$. Plotting the experimental data on a log-log scale reveals that at relatively high optical power, where we believe  the Kerr nonlinearity $\chi^{(3)}$ is saturated, the $E_N$-$R$ peak grows approximately linearly with $P_1$ and has a square root dependence on $P_2$. At lower power where the nonlinearity $\chi^{(3)}$ itself depends on the applied powers $P_1$ and $P_2$, the $E_N$-$R$ peak grows faster, and we find that the dependence on $P_1$ is approximately twice as steep.

The $E_N$-$R$ peak disappears if the laser is tuned to the $5S_{1/2}(F_g=2)\rightarrow5P_{1/2}(F_e=3)$ transition as shown in Fig. 3a.  As the mode matching condition is still satisfied this means that the Kerr nonlinearity $\chi^{(3)}$ required for efficient wave mixing is not enhanced in the proximity of the Raman resonance. It has been shown analytically that non-absorbing coherent states cannot be prepared on a transition $F_g \rightarrow F_e$ if $F_g < F_e$ \cite{SMIRNOV89}. Given the high degree of the beat signal suppression (~10$^{3}$) for the $F_g=2\rightarrow F_e=3$ transition, the present observation can be considered as high-fidelity experimental proof of this theoretical result. The non-existence of the $E_N$-$R$ peak is also interesting in the context of \emph{anomalous}  electromagnetically induced absorption (EIA) \cite{anomEIA}.  Experimentally found conditions for conventional EIA \cite{LEZAMA99} have been supported by analytical consideration and numerical modelling assuming an arbitrarily intense drive field and a weak probe \cite{TAICH00,GOREN03}. However, narrow EIA-type absorption resonances have since been reported   in systems that do not completely satisfy these conditions \cite{Korea03,ZAS-EIA}. Conversion of EIT to EIA and vice versa are under  investigation by a number of groups \cite{KUHN08, Brazh11, Auz12}, and application of the wave mixing method in combination with the drive-probe method could help in understanding the observed effects.

The extremely high spectral selectivity arising from the high temporal coherence of the new spectral components is another nice feature of the wave mixing technique. Beat signal peaks separated by $\delta\simeq$  200 kHz are well resolved despite the laser linewidth being approximately 500 kHz.  With mutually coherent optical fields the ultimate width of the beat signals is limited not by the laser linewidth but by frequency and amplitude noise introduced by the AOMs.  Figure 3b demonstrates sub-kHz resolution of the coherent heterodyne method.

Parallel polarizations are seldom used for probing coherent atomic media with the drive-probe method, primarily because of the technical issue of separating the co-propagating probe and drive fields. This issue is not a matter of concern for the coherent heterodyne method.  Figure 4 shows that the $E_N$-$R$ peak on the $5S_{1/2}(F_g=2)\rightarrow5P_{1/2}(F_e =3)$ transition for parallel polarizations is as strong as on the $5S_{1/2}(F_g=3)\rightarrow5P_{1/2}(F_e =2)$ transition, in contrast to the case of orthogonal polarizations. Both CPO and ground-state coherence contribute to the new field generation on the $5S_{1/2}(F_g=3)\rightarrow5P_{1/2}(F_e =2)$ transition.  The $E_N$-$R$  beat signal obtained on the $5S_{1/2}(F_g=2)\rightarrow5P_{1/2}(F_e =3)$ transition for parallel polarizations we attribute entirely to CPO because, as discussed above, a non-absorbing coherent state cannot be prepared on this transition.  For parallel linear polarizations of the applied fields the atom-light interaction could be described as a sequence of transitions between single magnetic sublevels in the ground state and single magnetic sublevels in the excited $5S_{1/2}(F_g=3)$ level, i.e.  $(\pi-\pi)$-,  $(\sigma^+-\sigma^+)$-  or $(\sigma^--\sigma^-)$- type transitions, at frequencies $\nu_1$ and $\nu_2$.  Such a combination of transitions cannot generate coherence between different ground-state magnetic sublevels, but it can result in population oscillations between the ground and excited levels.

\begin{figure}[b] 
\includegraphics[angle=0, width=10cm]{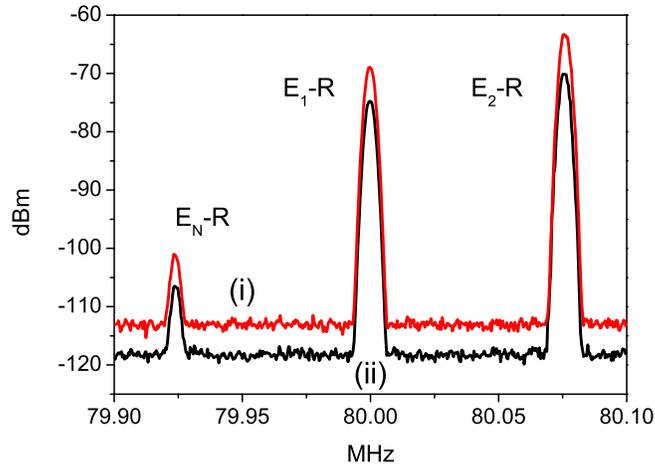}
\caption{RF beat signal for parallel linear polarizations of the applied radiation resonant with (i) the $^{85}$Rb $5S_{1/2}(F_g=3)\rightarrow5P_{1/2}(F_e=2)$ and (ii) the $5S_{1/2}(F_g=2)\rightarrow5P_{1/2}(F_e=3)$ transitions. Curve (i) is shifted up on 5 dBm.
   \label{fig.4}
   }
\end{figure}

\begin{figure}[t] 
\includegraphics[angle=0, width=10cm]{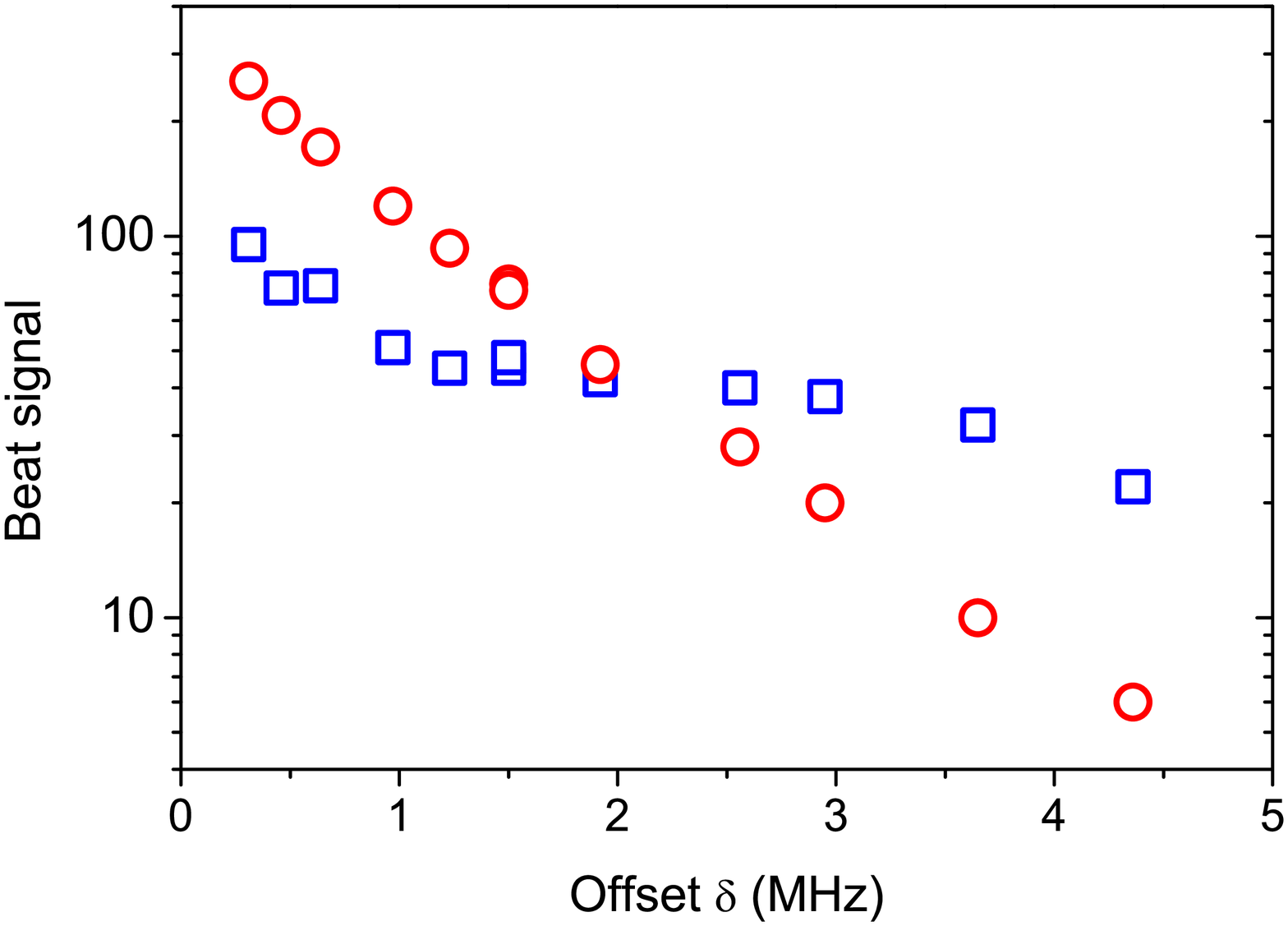}
\caption{Beat signal as a function of the frequency offset $\delta$ for parallel (blue squares) and orthogonal (red circles) polarizations obtained on the $^{85}$Rb $5S_{1/2}(F_g=3)\rightarrow5P_{1/2}(F_e=2)$ transition.
   \label{fig.5}
   }
\end{figure}

CPO leads to an oscillating atomic susceptibility $\chi$ at $\delta = \nu_{2}- \nu_{1}$.  The oscillating imaginary part of $\chi$ in turn results in phase modulation of the transmitted light and hence in optical sideband generation \cite{AKU-JPB-11}.  
The $E_N$-$R$ peak at $2\nu_1-\nu_2=\nu_1-\delta$ is the result of mixing the reference radiation with the low frequency sideband of the optical field at $\nu_1$.  The beats of the high frequency sideband at $\nu_1+\delta$ with the reference field spectrally coincides with the $E_2$-$R$ beat signal.

As CPO and CPT exhibit quite different spectral properties, the variation of the $E_N$-$R$ beat signal with frequency offset $\delta$ can also be used to distinguish the two processes.  Figure 5 shows the dependence on $\delta$ of the amplitude of the $E_N$-$R$ beat signal obtained on the $^{85}$Rb $5S_{1/2}(F_g=3)\rightarrow5P_{1/2}(F_e=2)$ for parallel and orthogonal polarizations.  Total variations of the $E_N$-$R$ peak in the case of orthogonal polarizations, when CPT plays a key role, are at least an order of magnitude larger compared to the parallel orientation.

The spectral dependence of nearly degenerate FWM enhanced by ground-state Zeeman coherence should reflect the spectral range over which the Kerr nonlinearity $\chi^{(3)}$ is significant.  It was demonstrated in \cite{AKU-JOB-04} that the absolute value of $\chi^{(3)}$ reaches a maximum when $\delta \simeq\gamma/2$, where $\gamma$ is the FWHM of the CPT resonance, then steadily decreases for larger values of $\delta$. At small $\delta$ ($\delta <\gamma/2$) reduced values of the Kerr nonlinearity are compensated by enhanced transmission of the applied radiation at $\nu_1$ and $\nu_2$ resulting in higher overall mixing efficiency.

By contrast, the characteristic spectral range in which the CPO mechanism operates  is determined by the natural width of the optical transition, $\Gamma/2\pi\simeq$ 6 MHz, which is much larger than a typical $\gamma\simeq$ 1 MHz in our experiment.  Thus, the beat signal for parallel polarizations has a rather flat dependence on $\delta$ compared to the rapid fall-off of the FWM-related beat signal that relies on the enhanced Kerr nonlinearity associated with CPT. The signal at $\delta  <$ 1 MHz, while still weaker compared to the orthogonal polarization case, exhibits a steep rise as $\delta$ approaches zero. The origin of this increase remains an open question  and requires additional investigation.

The results obtained on the Rb D1 line illustrate that new optical field generation can be efficient despite the depopulation of the resonant velocity group occurring on open transitions. Analyzing the atomic density dependence of the new spectral components we find that at present experimental conditions the low-frequency $E_N$-$R$ beat signal grows sharply with $N$, reaches a maximum at $N\simeq3\times10^{11}~cm^{-3}$ and then declines.


In conclusion, in our experiments performed on the Rb D1 line we have demonstrated that the spectrum of resonant laser light may be substantially altered as it propagates through an atomic medium. Both coherent population trapping and coherent population oscillations can contribute to this change. Despite their very similar appearance, the effects of CPT and CPO can be distinguished using a wave mixing technique by an appropriate choice of optical transition, frequency detuning between the optical fields and their polarizations.

A high degree of temporal coherence of new optical fields with the applied radiation allows the sub-kHz resolution and, therefore, the straightforward detection of spectral detail normally hidden by the laser linewidth. With high fidelity we have verified that dark coherent superposition responsible for electromagnetically induced transparency on the  $^{85}Rb$  $5S_{1/2}(F_g=2)\rightarrow5P_{1/2}(F_e=3)$ transition does not exist.

The resonant wave mixing approach with high detection capability not easily achievable with the well-established drive-probe method appears to be an efficient tool for investigating high-order atomic nonlinearities \cite{AKU-JOB-04}. The ability to distinguish the contributions to the nonlinear susceptibility made by long-lived Zeeman coherence and coherent population oscillations promises to be useful in designing atomic media with predetermined properties and understanding effects such as anomalous EIA. The method can be applied not only to alkali atoms, but to all atoms having hyperfine and magnetic structure in the ground state. Finally, the polarization flexibility, spatial selectivity and enhanced sensitivity compared to the drive-probe method could be useful for testing dilute laser-cooled atomic samples.

\bibliography{Hetero-D1}%

\end{document}